\documentclass{stacs_proc}

\newcounter{rivinro}

\newcommand{\re}{\addtocounter{rivinro}{1}\therivinro~~&}
\newcommand{\rf}{~~&}
\newcommand{\rs}{\addtocounter{rivinro}{1}\\\therivinro~~&}
\newcommand{\rt}{\\~~&}
\newcommand{\kohde}[1]{\newcounter{#1}\setcounter{#1}{\therivinro}}
\newenvironment{ohjelma}{%
\setcounter{rivinro}{0}%
\renewcommand{\>}{\hspace*{5mm}}%
\begin{tabular}{rl}}
{\end{tabular}}

\newcommand{\tb}[1]{\textbf{#1}}
\newcommand{\mc}[1]{\mathcal{#1}}
\newcommand{\var}[1]{\mathit{#1}}

\newcommand{\pp}{;\ ~}

\begin{document}

\title{Efficient Minimization of DFAs with Partial Transition Functions}

\author[TTY]{A. Valmari}{Antti Valmari}
\address[TTY]{Tampere University of Technology, Institute of Software
Systems,
PO Box 553, FI-33101 Tampere, Finland}
\email{{Antti.Valmari,Petri.Lehtinen}@tut.fi}
\thanks{Petri Lehtinen was funded by Academy of Finland,
  project ALEA (210795).}

\author[TTY]{P. Lehtinen}{Petri Lehtinen}


\begin{abstract}\noindent
Let \emph{PT-DFA} mean a deterministic finite automaton whose transition
relation is a partial function.
We present an algorithm for minimizing a PT-DFA in $O(m \lg n)$ time and
$O(m+n+\alpha)$ memory, where $n$ is the number of states, $m$ is the number
of \emph{defined} transitions, and $\alpha$ is the size of the alphabet.
Time consumption does not depend on $\alpha$, because the $\alpha$ term arises
from an array that is accessed at random and never initialized.
It is not needed, if transitions are in a suitable order in the input.
The algorithm uses two instances of an array-based data structure for
maintaining a refinable partition.
Its operations are all amortized constant time.
One instance represents the classical blocks and the other a partition of
transitions.
Our measurements demonstrate the speed advantage of our algorithm on
PT-DFAs over an $O(\alpha n \lg n)$ time, $O(\alpha n)$ memory algorithm.
\end{abstract}

\maketitle

\stacsheading{2008}{645-656}{Bordeaux}
\firstpageno{645}

\vskip-0.3cm
\section{Introduction}

Minimization of a deterministic finite automaton (DFA) is a classic problem in
computer science.
Let $n$ be the number of states, $m$ the number of transitions and $\alpha$
the size of the alphabet of the DFA.
Hopcroft made a breakthrough in 1970 by presenting an algorithm that runs in
$O(n \lg n)$ time, treating $\alpha$ as a constant~\cite{Hop70}.
Gries made the dependence of the running time of the algorithm on $\alpha$
explicit, obtaining $O(\alpha n \lg n)$~\cite{Gri73}.
(Complexity is reported using the RAM machine model under the uniform cost
criterion~\cite[p.\ 12]{AHU74}.)

Our starting point was the paper by Knuutila in 2001, where he presented yet
another $O(\alpha n \lg n)$ algorithm, and remarked that some versions which
have been believed to run within this time bound actually fail to do
so~\cite{Knu01}.
Hopcroft's algorithm is based on using only the ``smaller'' half of some set
(known as \emph{block}) that has been split.
Knuutila demonstrated with an example that although the most well-known notion
of ``smaller'' automatically leads to $O( \alpha n \lg n )$, two other notions
that have been used may yield $\Omega(n^3)$ when $\alpha = \frac{1}{2}n$.
He also showed that this can be avoided by maintaining, for each symbol, the
set of those states in the block that have input transitions labelled by that
symbol.
According to~\cite{Gri73}, Hopcroft's original algorithm did so.
Some later authors have dropped this complication as unnecessary, although it
is necessary when the alternative notions of ``smaller'' are used.

Knuutila mentioned as future work whether his approach can be used to
develop an $O(m \lg n)$ algorithm for DFAs whose
transition functions are not necessarily total.
For brevity, we call them PT-DFAs.
With an ordinary DFA, $O(m \lg n)$ is the same as $O( \alpha n \lg n )$ as $m
= \alpha n$, but with a PT-DFA it may be much better.
We present such an algorithm in this paper.
We refined Knuutila's method of maintaining sets of states with relevant input
transitions into a full-fledged data structure for maintaining refinable
partitions.
Instead of maintaining those sets of states, our algorithm maintains the
corresponding sets of transitions.
Another instance of the structure maintains the blocks.

Knuutila seems to claim that such a PT-DFA algorithm arises from the
results in~\cite{PaT87}, where an $O(m \lg n)$ algorithm was described for
refining a partition against a relation.
However, there $\alpha = 1$, so the solved problem is not an immediate
generalisation of ours.
Extending the algorithm to $\alpha > 1$ is not trivial, as can be appreciated
from the extension in~\cite{Fer89}.
It discusses $O(m \lg n)$ without openly promising it.
Indeed, its analysis treats $\alpha$ as a constant.
It seems to us that its running time does have an $\alpha n$ term.

In Section~\ref{Abstr_Algo} we present an abstract minimization algorithm
that, unlike~\cite{Gri73,Knu01}, has been adapted to PT-DFAs and avoids
scanning the blocks and the alphabet in nested loops.
The latter is crucial for converting $\alpha n$ into $m$ in the complexity.
The question of what blocks are needed in further splitting, has led to
lengthy and sometimes unconvincing discussions in earlier literature.
Our correctness proof deals with this issue using the ``loop invariant''
paradigm advocated in~\cite{Gri81}.
Our loop invariant ``knows'' what blocks are needed.

Section~\ref{Ref_Part} presents an implementation of the refinable partition
data structure.
Its performance relies on a carefully chosen combination of simple low-level
programming details.

The implementation of the main part of the abstract algorithm is the topic
of Section~\ref{Block-splitt}.
The analysis of its time consumption is based on proving of two lines of the
code that, whenever the line is executed again for the same transition, the end
state of the transition resides in a block whose size is at most half the size
in the previous time.
The numbers of times the remaining lines are executed are then related to these
lines.

With a time bound as tight as ours, the order in which the transitions are
presented in the input becomes significant, since the $\Theta(m \lg m)$ time
that typical good sorting algorithms tend to take does not necessarily fit
$O(m \lg n)$.
We discuss this problem in Section~\ref{Sort_Trans}, and present a solution
that runs in $O(m)$ time but may use more memory, namely $O(m + \alpha)$.

Some measurements made with our implementations of Knuutila's and our
algorithm are shown in Section~\ref{MeaCon}.

\section{Abstract Algorithm}\label{Abstr_Algo}

\newcommand{\Block}{\mathit{Block}}

A \emph{PT-DFA} is a 5-tuple $\mc{D} = (Q, \Sigma, \delta, \hat q, F)$
such that $Q$ and $\Sigma$ are finite sets, $\hat q \in Q$, $F \subseteq Q$
and $\delta$ is explained below.
The elements of $Q$ are called \emph{states}, $\hat q$ is the \emph{initial
state}, and $F$ is the set of \emph{final states}.
The set $\Sigma$ is the \emph{alphabet}.
We have $\delta \subseteq Q \times \Sigma \times Q$, and $\delta$ satisfies
the condition that if $(q,a,q_1) \in \delta$ and $(q,a,q_2) \in \delta$, then
$q_1 = q_2$.
The elements of $\delta$ are \emph{transitions}.
In essence, $\delta$ is a partial function from $Q \times \Sigma$ to $Q$.
Therefore, if $(q,a,q') \in \delta$, we write $\delta(q,a) = q'$.
If $q \in Q$ and $a \in \Sigma$ but there is no $q'$ such that $(q,a,q') \in
\delta$, we write $\delta(q,a) = \bot$, where $\bot$ is some symbol satisfying
$\bot \notin Q$.
We will use $|\delta|$ as the
number of transitions, and this number may be much smaller than $|Q||\Sigma|$,
which is the number of transitions if $\delta$ is a full function.

\newcommand{\rbleoo}[1]{{-}#1{\rightarrow}}
\newcommand{\rble}[1]{\,\rbleoo{#1}\,}
By $q \rble{a_1 a_2 \cdots a_n} q'$ we denote that there is a path from
state $q$ to state $q'$ such that the labels along the path constitute the
word $a_1 a_2 \cdots a_n$.
That is, $q \rble{\varepsilon} q$ holds for every $q \in Q$, and $q \rble{a_1
a_2 \cdots a_n a_{n+1}} q'$ holds if and only if there is some $q'' \in Q$
such that $q \rble{a_1 a_2 \cdots a_n} q''$ and $\delta(q'',a_{n+1}) = q' \neq
\bot$.
The \emph{language} accepted by $\mc{D}$ is the set of words labelling the
paths from the initial state to final states, that is, $\mc{L}(\mc{D}) = \{\,
\sigma \in \Sigma^*\ |\ \exists q \in F: \hat q \rble{\sigma} q \,\}$.
We will also talk about the languages of individual states, that is,
$\mc{L}(q) = \{\, \sigma \in \Sigma^*$ $|$ $\exists q' \in F: q \rble{\sigma}
q' \,\}$.
Obviously $\mc{L}(\mc{D}) = \mc{L}(\hat q)$.

We say that a state is \emph{relevant}, if and only if either it is the
initial state, or it is reachable from the initial state and some final state
is reachable from it.
More precisely, $R = \{\hat q \} \cup \{\, q \in Q \ |\ \exists q' \in F:
\exists \sigma \in \Sigma^*: \exists \rho \in \Sigma^*: \hat q \rble{\sigma} q
\rble{\rho} q' \,\}$.
It is obvious that irrelevant states and their adjacent transitions may be
removed from a PT-DFA without affecting its language.
The initial state cannot be removed, because otherwise the result would
violate the condition $\hat q \in Q$ in the definition of a DFA.
The removal yields the PT-DFA $(R, \Sigma, \delta', \hat q, F')$, where
$\delta' = \delta \cap (R \times \Sigma \times R)$ and $F' = F \cap R$.

If no final state is reachable from the initial state, then $\mc{L}(\mc{D}) =
\emptyset$.
This is handled as a special case in our algorithm, because otherwise the
result might contain unnecessary transitions from the initial state to itself.
For this purpose, let empty\_DFA$(\Sigma)$ be $(\{x\}, \Sigma, \emptyset, x,
\emptyset)$, where $x$ is just any element.
Obviously empty\_DFA$(\Sigma)$ is the smallest PT-DFA with the alphabet
$\Sigma$ that accepts the empty language.

\newcommand{\nes}[1]{\delta_{#1} \neq \emptyset}
\begin{figure}[t]
\begin{ohjelma}
\re
$(Q, \Sigma, \delta, \hat q, F)$ := remove irrelevant states and transitions
from $(Q, \Sigma, \delta, \hat q, F)$\kohde{alkuA}\rs
\tb{if} $F = \emptyset$ \tb{then} \tb{return}
  empty\_DFA($\Sigma$)\kohde{alkuB}\rs
\tb{else}\rs
\>\tb{if} $Q = F$ \tb{then} $\mc{B}$ := $\{F\}$
  \tb{else} $\mc{B}$ := $\{ F,\ Q-F \}$\kohde{Balku}\rs
\>$\mc{U}$ := $\{\, (B,a) \ |\ B \in \mc{B} \wedge a \in \Sigma \wedge
  \nes{B,a} \,\}$\kohde{Ualku}\rs
\>\tb{while} $\mc{U} \neq \emptyset$ \tb{do}\kohde{paaalku}\rs
\>\>$(B,a)$ := any\_element\_of$(\mc{U})$\pp $\mc{U}$ := $\mc{U} -
\{(B,a)\}$\kohde{poistaU}\rs
\>\>\tb{for} $C \in \mc{B}$ such that $\exists q \in C: \delta(q,a) \in B$
  \tb{do}\kohde{Cfor}\rs
\>\>\>$C_1$ := $\{\, q \in C \ |\ \delta(q,a) \in B \,\}$\pp
      $C_2$ := $C - C_1$\kohde{jaaC}\rs
\>\>\>\tb{if} $C_2 \neq \emptyset$ \tb{then}\kohde{Ctesti}\rs
\>\>\>\>$\mc{B} := \mc{B} - \{C\}$\pp
	$\mc{B}$ := $\mc{B} \cup \{C_1, C_2\}$\kohde{Bmuutos}\rs
\>\>\>\>\tb{if} $|C_1| \leq |C_2|$ \tb{then}
	  $\var{small}$ := 1; $\var{big}$ := 2
	\tb{else} $\var{small}$ := 2; $\var{big}$ := 1\rs
\>\>\>\>$\mc{U}$ := $\mc{U} \cup \{\, (C_\var{small},b) \ |\
	\nes{C_\var{small},b}
	\wedge b \in \Sigma \,\}$\kohde{Upienempi}\rs
\>\>\>\>$\mc{U}$ := $\mc{U} \cup \{\, (C_\var{big},b) \ |\ \nes{C_\var{big},b}
	\wedge (C,b) \in \mc{U} \,\}$\kohde{Uisompi}\rs
\>\>\>\>$\mc{U}$ := $\mc{U} - (\,\{C\} \times \Sigma\,)
	$\kohde{Uvanha}\kohde{paaloppu}\rs
\>$Q'$ := $\mc{B}$\pp $\delta'$ := $\emptyset$\pp
  $\hat q'$ := $\Block(\hat q)$\pp $F'$ := $\emptyset$\kohde{tulosAlku}\rs
\>\tb{for} $B \in \mc{B}$ \tb{do}\rs
\>\>$q$ := any\_element\_of$(B)$\rs
\>\>\tb{if} $q \in F$ \tb{then} $F'$ := $F' \cup \{B\}$\kohde{Fpilkku}\rs
\>\>\tb{for} $a \in \Sigma$ such that $\delta(q,a) \neq \bot$ \tb{do}
  $\delta'$ := $\delta' \cup
  \{\:(B,a,\Block(\delta(q,a)))\:\}$\kohde{deltapilkku}\kohde{tulosLoppu}\rs
\>\tb{return} $(Q', \Sigma, \delta', \hat q', F')$
\end{ohjelma}
\caption{Abstract PT-DFA minimization algorithm}\label{abstr-algo}
\end{figure}

The abstract minimization algorithm is shown in Figure~\ref{abstr-algo}.
In it, $\mc{B}$ denotes a \emph{partition} on $Q$.
That is, $\mc{B}$ is a collection $\{ B_1, B_2, \ldots, B_n \}$ of nonempty
subsets of $Q$ such that $B_1 \cup B_2 \cup \cdots \cup B_n = Q$, and
$B_i \cap B_j = \emptyset$ whenever $1 \leq i < j \leq n$.
The elements of $\mc{B}$ are called \emph{blocks}.
By checking all statements that modify the contents of $\mc{B}$, it is easy to
verify that after its initialization on line~\theBalku, $\mc{B}$ is a
partition on $Q$ throughout the execution of the algorithm, except temporarily
in the middle of line~\theBmuutos.

By $\Block(q)$ we denote the block to which state $q$ belongs.
Therefore, if $q \in Q$, then $q \in \Block(q) \in \mc{B}$.
For convenience, we define $\Block(\bot) = \bot \notin \mc{B}$.
If $\Block(q_1) \neq \Block(q_2)$ ever starts to hold, then it stays valid up
to the end of the execution of the algorithm.

Elements of $\mc{B} \times \Sigma$ are called \emph{splitters}.
Let $\delta_{B,a} = \{\, (q,a,q') \in \delta \ |\ q' \in B \,\}$.
We say that splitter $(B,a)$ is \emph{nonempty}, if and only if $\nes{B,a}$.
The set $\mc{U}$ contains those nonempty splitters that are currently
``unprocessed''.
It is obvious from line~\theCfor\ that empty splitters would have no effect.
The main loop of the algorithm (lines \thepaaalku\ldots\thepaaloppu)
starts with all nonempty splitters as unprocessed, and ends when no nonempty
splitter is unprocessed.
The classic algorithm uses either only $F$ or only $Q-F$ for constructing the
initial splitters, but this does not work with a partial $\delta$.

The goal of the main loop is to split blocks until they are consistent with
$\delta$, without splitting too much.
We will now prove in two steps that this is achieved.

\begin{lemma}\label{ei-liikaa-lemma}
For every $q_1 \in Q$ and $q_2 \in Q$, if $\Block(q_1) \neq \Block(q_2)$ at
any time of the execution of the algorithm in Figure~\ref{abstr-algo}, then
$\mc{L}(q_1) \neq \mc{L}(q_2)$.
\end{lemma}
\proof
If the algorithm puts states $q_1$ and $q_2$ into different blocks on line
\theBalku, then either $\varepsilon \in \mc{L}(q_1) \wedge \varepsilon \notin
\mc{L}(q_2)$ or $\varepsilon \notin \mc{L}(q_1) \wedge \varepsilon \in
\mc{L}(q_2)$.
Otherwise, it does so on line~\theBmuutos.
Then there are $i$, $j$, $B$ and $a$ such that $\{i,j\} = \{1,2\}$,
$\delta(q_i,a) \in B$ and $\delta(q_j,a) \notin B$.
Let $q'_i = \delta(q_{i},a)$.

If $\delta(q_{j},a) \neq \bot$, then let $q'_j = \delta(q_{j},a)$.
We have $q'_j \notin B$.
Because the algorithm has already put $q'_i$ and $q'_j$ into different blocks
(they were in different blocks on line~9),
there is some $\sigma \in \Sigma^*$ such that either $\sigma \in \mc{L}(q'_i)
\wedge \sigma \notin \mc{L}(q'_j)$ or vice versa.
As a consequence, $a\sigma$ is in $\mc{L}(q_1)$ or in $\mc{L}(q_2)$, but not
in both.

Assume now that $\delta(q_{j},a) = \bot$.
Because of lines \thealkuA\ and \thealkuB, $\mc{L}(q) \neq \emptyset$ for every
$q \in Q$.
There is thus some $\sigma \in \Sigma^*$ such that $\sigma \in \mc{L}(q'_i)$.
We have $a\sigma \in \mc{L}(q_{i})$.
Clearly $a\sigma \notin \mc{L}(q_{j})$.
\qed

At this point it is worth noticing that line~\thealkuA\ is important for the
correctness of the algorithm.
Without it, there could be two reachable states $q_1$ and $q_2$ that accept the
same language, and $a$ such that $\delta(q_1,a) = \bot$ while $\delta(q_2,a)$
is a state that accepts the empty language.
The algorithm would eventually put $q_1$ and $q_2$ into different blocks.

We have shown that the main loop does not split blocks when it should not.
We now prove that it splits all the blocks that it should.

\begin{lemma}\label{tarpeeksi-lemma}
At the end of the algorithm in Figure~\ref{abstr-algo}, for every $q_1 \in Q$,
$q_2 \in Q$ and $a \in \Sigma$, if $\Block(q_1) = \Block(q_2)$, then
$\Block(\delta(q_1,a)) = \Block(\delta(q_2,a))$.
\end{lemma}
\proof
To improve readability, let $B_1 = \Block(\delta(q_1,a))$ and $B_2 =
\Block(\delta(q_2,a))$.
In the proof, $\Block()$, $B_1$ and $B_2$ are always evaluated with the
current $\mc{B}$, so their contents change.
The proof is based on the following loop invariant:
\begin{quote}
On line~\thepaaalku, for every $q_1 \in Q$, $q_2 \in Q$ and $a \in \Sigma$,
if $\Block(q_1) = \Block(q_2)$, then $B_1 = B_2$ or $(B_1,a) \in \mc{U}$ or
$(B_2,a) \in \mc{U}$.
\end{quote}

Consider the situation immediately after line~\theUalku.
If $B_1 \neq \bot$, then $(B_1,a) \in \mc{U}$.
If $B_2 \neq \bot$, then $(B_2,a) \in \mc{U}$.
If $B_1 = B_2 = \bot$, then $B_1 = B_2$.
Thus the invariant holds initially.

Consider any $q_1$, $q_2$, $a$ and instance of executing line~\thepaaalku\
such that the invariant holds.
Our task is to show that the invariant holds for them also when
line~\thepaaalku\ is executed for the next time.

The case that the invariant holds because $\Block(q_1) \neq \Block(q_2)$ is
simple.
Blocks are never merged, so $\Block(q_1) \neq \Block(q_2)$ is valid also the
next time.

Consider the case $\Block(q_1) = \Block(q_2)$, $B_1 \neq B_2$ and $(B_i,a) \in
\mc{U}$, where $i = 1$ or $i = 2$.
Let $j = 3-i$.
If $(B_i,a)$ is the $(B,a)$ of line~\thepoistaU, then, when $\Block(q_i)$ is
the $C$ of the \tb{for}-loop, $q_{i}$ goes to $C_1$ and $q_{j}$ goes to $C_2$.
So $\Block(q_1) = \Block(q_2)$ ceases to hold, rescuing the invariant.
If $(B_i,a)$ is not the $(B,a)$ of line~\thepoistaU, then, whenever $B_i$ is
split, lines~\theUpienempi\ and~\theUisompi\ take care that both halves end up
in $\mc{U}$.
Thus $(B_i,a) \in \mc{U}$ stays true keeping the invariant valid, although
$B_i$ and $\mc{U}$ may change.

Let now $\Block(q_1) = \Block(q_2)$ and $B_1 = B_2$.
To invalidate the invariant, $B_1$ or $B_2$ must be changed so that $B_1 =
B_2$ ceases to hold.
When this happens, line~\theUpienempi\ puts $(B_i,a)$ into $\mc{U}$, where $i
= 1$ or $i = 2$.
Like above, lines~\theUpienempi\ and~\theUisompi\ keep $(B_i,a)$ in $\mc{U}$
although $B_i$ may change until line~\thepaaalku\ is entered again.

We have completed the proof that the invariant stays valid.

When line~\thetulosAlku\ is entered, $\mc{U} = \emptyset$.
The invariant now yields that 
if $\Block(q_1) = \Block(q_2)$, then $\Block(\delta(q_1,a)) =
\Block(\delta(q_2,a))$.
\qed

It is not difficult to check that lines~\thetulosAlku\ldots\thetulosLoppu\
yield a PT-DFA, that is, $Q'$ and $\Sigma$ are finite sets and so on.
In particular, the construction gives $\delta'(B,a)$ a value at most once.
We now show that the result is the right PT-DFA.

\begin{lemma}
Let $\mc{D}' = (Q', \Sigma, \delta', \hat q', F')$ be the result of the
algorithm in Figure~\ref{abstr-algo}.
We have $\mc{L}(\mc{D}') = \mc{L}(\mc{D})$.
Furthermore, every PT-DFA that accepts $\mc{L}(\mc{D})$ has at least as
many states and transitions as $\mc{D}'$.
If it has the same number of states, it is either isomorphic with $\mc{D}'$
(ignoring $\Sigma$ in the comparison),
or it is of the form $(\{\hat q''\}, \Sigma'', \delta'', \hat q'', \emptyset)$
with $\delta'' \neq \emptyset$.
\end{lemma}
\proof
The case where the algorithm exits on line~\thealkuB\ is trivial and has been
discussed, so from now on we discuss the case where the algorithm goes through
the main part.

Let $q \in Q$ and $a \in \Sigma$.
Lemma~\ref{tarpeeksi-lemma} implies that $\Block(\delta(q,a)) =
\Block(\delta(q',a))$ for every $q' \in \Block(q)$.
From this line~\thedeltapilkku\ yields $\delta'(\Block(q),a) =
\Block(\delta(q,a))$.
By induction, if $\sigma \in \Sigma^*$, $q' \in Q$ and $q \rble{\sigma} q'$ in
$\mc{D}$ then $\Block(q) \rble{\sigma} \Block(q')$ in $\mc{D}'$, and if
$\Block(q) \rble{\sigma} B \neq \bot$ in $\mc{D}'$ then there is $q' \in Q$
such that $B = \Block(q')$ and $q \rble{\sigma} q'$ in $\mc{D}$.
Similarly, lines~\theBalku\ and \theFpilkku\ guarantee that $q' \in F$ if and
only if $\Block(q') \in F'$.
Together these yield $\mc{L}(q) = \mc{L}(\Block(q))$ and, in particular,
$\mc{L}(\mc{D}) = \mc{L}(\hat q) = \mc{L}(\hat q') = \mc{L}(\mc{D}')$.

Let $(Q'', \Sigma, \delta'', \hat q'', F'')$ be any PT-DFA that accepts
the same language as $\mc{D}'$.
Let $q' \in Q'$.
Because the algorithm executed the main part, there are some $\sigma \in
\Sigma^*$ and $\rho \in \Sigma^*$ such that $\hat q' \rble{\sigma} q'$ and
$\rho \in \mc{L}(q')$.
So $\sigma\rho \in \mc{L}(\hat q') = \mc{L}(\hat q'')$, and also $Q''$
contains a state $q''$ such that $\hat q'' \rble{\sigma} q''$ and $\mc{L}(q'')
= \mc{L}(q')$.
As $\sigma$ may vary, there may be many $q''$ with $\mc{L}(q'') = \mc{L}(q')$.
We arbitrarily choose one of them and denote it with $f(q')$.
Lemma~\ref{ei-liikaa-lemma} implies that if $q'_1 \neq q'_2$, then
$\mc{L}(q'_1) \neq \mc{L}(q'_2)$, yielding $f(q'_1) \neq f(q'_2)$.
So $|Q''| \geq |Q'|$.
If $\delta'(q',a) \neq \bot$, then some $a \rho' \in \mc{L}(q') =
\mc{L}(f(q'))$, so $\delta''(f(q'),a) \neq \bot$.
As a consequence, $|\delta''| \geq |\delta'|$.

If $|Q''| = |Q'|$, then $f$ is an isomorphism.
\qed

The proof has the consequence that after the end of the main loop,
$\Block(q_1)$ $=$ $\Block(q_2)$ if and only if $\mc{L}(q_1) = \mc{L}(q_2)$.

Let us consider the number of times a transition $(q,a,q')$ can be used on
line~\thejaaC.
It is used whenever such a $(B,a)$ is taken from $\mc{U}$ that $q' \in B$,
that is, $\Block(q') = B$.
So, shortly before using $(q,a,q')$, $(\Block(q'),a) \in \mc{U}$ held but
ceased to hold (line~\thepoistaU).
To use it again, $(\Block(q'),a) \in \mc{U}$ must be made to hold again.
To make $(\Block(q'),a) \in \mc{U}$ to hold again, line~\theUpienempi\
or~\theUisompi\ must be executed such that $\Block(q')$ is in the role of
$C_\var{small}$ or $C_\var{big}$, and $a$ is in the role of $b$.
But line~\theUisompi\ tests that $(C,b) \in \mc{U}$, so it cannot make
$(\Block(q'),a) \in \mc{U}$ to hold if it did not hold already on line~9,
although it can keep $(\Block(q'),a) \in \mc{U}$ valid.
So only line~\theUpienempi\ can make $(\Block(q'),a) \in \mc{U}$ to hold
again.
An important detail of the algorithm is that line~\theUpienempi\ puts the
\emph{smaller} half of $C$ (paired with $a$) into $\mc{U}$.
Therefore, each time $(\Block(q'),a) \in \mc{U}$ starts to hold again, $q'$
resides in a block whose size is at most half of the size in the previous time.
As a consequence, $(q,a,q')$ can be used for splitting at most $\lg |Q|+1$
times.

\section{Refinable Partitions}\label{Ref_Part}

\newcommand{\False}{\textsf{False}}
\newcommand{\True}{\textsf{True}}
\newcommand{\Nil}{\textsf{Nil}}

\newcommand{\func}[1]{\mathit{#1}}

\newcommand{\Empty}{\func{Empty}}
\newcommand{\Add}{\func{Add}}
\newcommand{\Remove}{\func{Remove}}
\newcommand{\Size}{\func{Size}}
\newcommand{\Set}{\func{Set}}
\newcommand{\First}{\func{First}}
\newcommand{\Next}{\func{Next}}
\newcommand{\Mark}{\func{Mark}}
\newcommand{\Split}{\func{Split}}
\newcommand{\NoMarks}{\func{No\_marks}}

\newcommand{\elems}{\var{elems}}
\newcommand{\loc}{\var{loc}}
\newcommand{\sidx}{\var{sidx}}
\newcommand{\sfirst}{\var{first}}
\newcommand{\send}{\var{end}}
\newcommand{\smid}{\var{mid}}
\newcommand{\sets}{\var{sets}}

The refinable partition data structure maintains a partition of the set $\{1,
\ldots, \var{max}\}$.
Our algorithm uses one instance of it with $\var{max} = |Q|$ for the blocks
and another with $\var{max} = |\delta|$ for the splitters.
Each set in the partition has an index in the range $1, \ldots, \sets$, where
$\sets$ is the current number of sets.
The structure supports the following operations.

\begin{description}

\item[$\Size(s)$]
Returns the number of elements in the set with index $s$.

\item[$\Set(e)$]
Returns the index of the set that element $e$ belongs to.

\item[$\First(s)$ and $\Next(e)$]
The elements of the set $s$ can be scanned by executing first
$e$~:=~$\First(s)$ and then \tb{while} $e \neq 0$ \tb{do} $e$ := $\Next(e)$.
Each element will be returned exactly once, but the ordering in which they are
returned is unspecified.
While scanning a set, $\Mark$ and $\Split$ must not be executed.

\item[$\Mark(e)$]
Marks the element $e$ for splitting of a set.

\item[$\Split(s)$]
If either none or all elements of set $s$ have been marked, returns $0$.
Otherwise removes the marked elements from the set, makes a new set of the
marked elements, and returns its index.
In both cases, unmarks all the elements in the set or sets.

\item[$\NoMarks(s)$]
Returns $\True$ if and only if none of the elements of $s$ is marked.

\end{description}
The implementation uses the following $\var{max}$-element arrays.

\begin{description}

\item[$\elems$]
Contains $1, \ldots, \var{max}$ in such an order that elements that belong to
the same set are next to each other.

\item[$\loc$]
Tells the location of each element in $\elems$, that is, $\elems[ \loc[e]] =
e$.

\item[$\sidx$]
The index of the set that $e$ belongs to is $\sidx[e]$.

\item[$\sfirst$ and $\send$]
The elements of set $s$ are $\elems[f]$, $\elems[f+1]$, \ldots,
$\elems[\ell]$, where $f = \sfirst[s]$ and $\ell = \send[s] - 1$.

\item[$\smid$]
Let $f$ and $\ell$ be as above, and let $m = \smid[s]$.
The marked elements are $\elems[f]$, \ldots, $\elems[m-1]$, and the unmarked
are $\elems[m]$, \ldots, $\elems[\ell]$.

\end{description}
Initially $\sets = 1$, $\sfirst[1] = \smid[1] = 1$, $\send[1] = \var{max}+1$,
and $\elems[e] = \loc[e] = e$ and $\sidx[e] = 1$ for $e \in \{1, \ldots,
\var{max}\}$.
Initialization takes $O(\var{max})$ time and $O(1)$ additional memory.

The implementation of the operations is shown in Figure~\ref{ref-part-ds}.
Each operation runs in constant time, except $\Split$, whose worst-case time
consumption is linear in the number $M$ of marked elements.
However, also $\Split$ can be treated as constant-time in the analysis of our
algorithm, because it is amortized constant time.
When calling $\Split$, there had been $M$ calls of $\Mark$.
They are unique to this call of $\Split$, because $\Split$ unmarks the
elements in question.
The total time consumption of these calls of $\Mark$ and $\Split$ is
$\Theta(M)$, but the same result is obtained even if $\Split$ is treated as
constant-time.

\newcommand{\kork}{\rule{0pt}{15pt}}
\begin{figure}[t]
\begin{ohjelma}\rf
$\underline{\Size(s)}$\rt
\tb{return} $\send[s] - \sfirst[s]$
\rt\kork
$\underline{\Set(e)}$\rt
\tb{return} $\sidx[e]$
\rt\kork
$\underline{\First(s)}$\rt
\tb{return} $\elems[\sfirst[s]]$
\rt\kork
$\underline{\Next(e)}$\rt
\tb{if} $\loc[e]+1 \geq \send[\sidx[e]]$ \tb{then} \tb{return} $0$\rt
\tb{else} \tb{return} $\elems[\loc[e]+1]$
\rt\kork
$\underline{\Mark(e)}$\rt
$s$ := $\sidx[e]$\pp $\ell$ := $\loc[e]$\pp $m$ := $\smid[s]$\rt
\tb{if} $\ell \geq m$ \tb{then}\rt
\>$\elems[\ell]$ := $\elems[m]$\pp $\loc[\elems[\ell]]$ := $\ell$\rt
\>$\elems[m]$ := $e$\pp $\loc[e]$ := $m$\pp $\smid[s]$ := $m + 1$
\rt\kork
$\underline{\Split(s)}$\rt
\tb{if} $\smid[s] = \send[s]$ \tb{then} $\smid[s] := \sfirst[s]$\rt
\tb{if} $\smid[s] = \sfirst[s]$ \tb{then} \tb{return} $0$\rt
\tb{else}\rt
\>$\sets$ := $\sets + 1$\rt
\>$\sfirst[\sets]$ := $\sfirst[s]$\pp $\smid[\sets]$ := $\sfirst[s]$\pp
  $\send[\sets]$ := $\smid[s]$\rt
\>$\sfirst[s]$ := $\smid[s]$\rt
\>\tb{for} $\ell$ := $\sfirst[\sets]$ \tb{to} $\send[\sets] - 1$ \tb{do}
  $\sidx[\elems[\ell]]$ := $\sets$\rt
\>\tb{return} $\sets$
\rt\kork
$\underline{\NoMarks(s)}$\rt
\tb{if} $\smid[s] = \sfirst[s]$ \tb{then} \tb{return} $\True$\rt
\tb{else} \tb{return} $\False$
\end{ohjelma}
\caption{Implementation of the refinable partition data structure}%
\label{ref-part-ds}
\end{figure}

\newcommand{\BRP}{\var{BRP}}
\newcommand{\TRP}{\var{TRP}}
\newcommand{\tail}{\var{tail}}
\newcommand{\name}{\var{name}}
\newcommand{\head}{\var{head}}
\newcommand{\InTrs}{\var{In\_trs}}
\newcommand{\URSpls}{\var{Unready\_Spls}}
\newcommand{\BTouched}{\var{Touched\_Blocks}}
\newcommand{\UTouched}{\var{Touched\_Spls}}

\vskip-0.3cm
\section{Block-splitting Stage}\label{Block-splitt}

In this section we show how lines~\theBalku\ldots\thepaaloppu\ of the abstract
algorithm can be implemented in $O(|\delta| \lg |Q|)$ time and $O(
|\delta| )$ memory assuming that $\delta$ is available in a suitable
ordering.
The implementation of abstract lines 1\ldots3 and 16\ldots21 in $O(|Q| +
|\delta|)$ time and memory is easy and not discussed further in this paper.
(By ``abstract lines'' we refer to lines in Figure~\ref{abstr-algo}).

The implementation relies on the following data structures.
The ``simple sets'' among them are all initially empty.
They have only three operations, all $O(1)$ time:
the set is empty if and only if $\Empty$ returns $\True$, 
$\Add(i)$ adds number $i$ to the set without checking if it already is there,
and $\Remove$ removes any number from the set and returns the removed number.
The implementation may choose freely the element that $\Remove$ removes and
returns.
One possible efficient implementation of a simple set consists of an array
that is used as a stack.

\begin{description}

\item[$\tail$, $\var{label}$ and $\var{head}$]
The transitions have the indices $1$, \ldots, $|\delta|$.
If $t$ is the index of the transition $(q,a,q')$, then $\tail[t] = q$,
$\var{label}[t] = a$, and $\var{head}[t] = q'$.

\item[$\InTrs$]
This stores the indices of the input transitions of state $q$.
The ordering of the transitions does not matter.
This is easy to implement efficiently.
For instance, one may use an array $\elems$ of size $|\delta|$, together with
arrays $\sfirst$ and $\send$ of size $|Q|$, so that the indices of the input
transitions of $q$ are $\elems[\sfirst[q]]$, $\elems[\sfirst[q]+1]$, \ldots,
$\elems[\send[q]-1]$.
The array can be initialized in $O(|Q| + |\delta|)$ time with counting sort,
using $\var{head}[t]$ as the key.

\item[$\BRP$]
This is a refinable partition data structure on $\{1, \ldots, |Q|\}$.
It represents $\mc{B}$, that is, the blocks.
The index of the set in $\BRP$ is used as the index of the block also
elsewhere in the algorithm.
Initially $\BRP$ consists of one set that contains the indices of the
states.

\item[$\TRP$]
This is a refinable partition data structure on $\{1, \ldots, |\delta|\}$.
Each of the sets in it consists of the indices of the input transitions of
some nonempty splitter $(B,a)$.
That is, $\TRP$ stores $\{\, \delta_{B,a} \ |\ B \in \mc{B} \wedge a
\in \Sigma \wedge \delta_{B,a} \neq \emptyset \,\}$.
The index of $\delta_{B,a}$ in $\TRP$ is used as the index of $(B,a)$ also
elsewhere in the algorithm.
For this reason, we will occasionally use the word ``splitter'' also of the
sets in $\TRP$.
Initially $\TRP$ consists of $\{\, \delta_{Q,a} \ |\ a \in \Sigma \wedge
\delta_{Q,a} \neq \emptyset \,\}$, that is, two transitions are in the same
set if and only if they have the same label.
This can be established as follows:
\begin{center}
\begin{ohjelma}\rf
\tb{for} $a \in \Sigma$ such that $\delta_{Q,a} \neq \emptyset$ \tb{do}\rt
\>\tb{for} $t \in \delta_{Q,a}$ \tb{do} $\TRP.\Mark(t)$\rt
\>$\TRP.\Split(1)$
\end{ohjelma}
\end{center}
If transitions are pre-sorted such that transitions with the same label are
next to each other, then this runs in $O(|\delta|)$ time and $O(1)$ additional
memory.

\item[$\URSpls$]
This is a simple set of numbers in the range $1$, \ldots, $|\delta|$.
It stores the indices of the unprocessed nonempty splitters.
That is, it implements the $\mc{U}$ of the abstract algorithm.
Because each nonempty splitter has at least one incoming transition and
splitters do not share transitions, $|\delta|$ suffices for the range.

\item[$\BTouched$]
This is a simple set of numbers in the range $1$, \ldots, $|Q|$.
It contains the indices of the blocks $C$ that were met when
backwards-traversing the incoming transitions of the current splitter on
abstract line~8.
It is always empty on line~19.

\item[$\UTouched$]
This is a simple set of numbers in the range $1$, \ldots, $|\delta|$.
It contains the indices of the splitters that were affected when
scanning the incoming transitions of the smaller of the new blocks that
resulted from a split.
It is empty on line~4.

\end{description}

\newcommand{\SplitBlock}{\func{Split\_block}}
\newcommand{\MainPart}{\func{Main\_part}}
\begin{figure}[t]
\begin{ohjelma}

&\underline{$\SplitBlock(b)$}\rs
$b'$ := $\BRP.\Split(b)$\rs
\tb{if} $b' \neq 0$ \tb{then}\rs
\>\tb{if} $\BRP.\Size(b) < \BRP.\Size(b')$ \tb{then} $b'$ := $b$\rs
\>$q$ := $\BRP.\First(b')$\rs
\>\tb{while} $q \neq 0$ \tb{do}\rs
\>\>\tb{for} $t \in \InTrs[q]$ \tb{do}\rs
\>\>\>$p$ := $\TRP.\Set(t)$\rs
\>\>\>\tb{if} $\TRP.\NoMarks(p)$ \tb{then} $\UTouched.\Add(p)$\rs
\>\>\>$\TRP.\Mark(t)$\rs
\>\>$q$ := $\BRP.\Next(q)$\rs
\>\tb{while} $\neg \UTouched.\Empty$ \tb{do}\rs
\>\>$p$ := $\UTouched.\Remove$\rs
\>\>$p'$ := $\TRP.\Split(p)$\rs
\>\>\tb{if} $p' \neq 0$ \tb{then} $\URSpls.\Add(p')$\\[2mm]

&\underline{$\MainPart$}\rs
Initialize $\TRP$ to $\{\, \delta_{Q,a} \ |\ a \in \Sigma \wedge \delta_{Q,a}
\neq \emptyset \,\}$\rs
\tb{for} $p$ := $1$ \tb{to} $\TRP.\sets$ \tb{do} $\URSpls.\Add(p)$\rs
\tb{for} $q \in F$ \tb{do} $\BRP.\Mark(q)$\rs
$\SplitBlock(1)$\rs
\tb{while} $\neg \URSpls.\Empty$ \tb{do}\kohde{alku}\rs
\>$p$ := $\URSpls.\Remove$\rs
\>$t$ := $\TRP.\First(p)$\rs
\>\tb{while} $t \neq 0$ \tb{do}\rs
\>\>$q$ := $\tail[t]$\pp $b'$ := $\BRP.\Set(q)$\rs
\>\>\tb{if} $\BRP.\NoMarks(b')$ \tb{then} $\BTouched.\Add(b')$\rs
\>\>$\BRP.\Mark(q)$\rs
\>\>$t$ := $\TRP.\Next(t)$\rs
\>\tb{while} $\neg \BTouched.\Empty$ \tb{do}\rs
\>\>$b$ := $\BTouched.\Remove$\rs
\>\>$\SplitBlock(b)$
\end{ohjelma}
\caption{Implementation of lines~\theBalku\ldots\thepaaloppu\ of the abstract
  algorithm}\label{concr-algo}
\end{figure}
The block-splitting stage is shown in Figure~\ref{concr-algo}.
We explain its operation in the proof of the following theorem.
\begin{theorem}
Given a PT-DFA all whose states are relevant and that has at least one
final state, the algorithm in Figure~\ref{concr-algo} computes the same
$\mc{B}$ (represented by $\BRP$) as lines~4\ldots15 of
Figure~\ref{abstr-algo}.
\end{theorem}
\proof
Let us first investigate the operation of $\SplitBlock$.
As was told earlier, $\BRP$ models $\mc{B}$, $\TRP$ models the set of all
nonempty splitters (or the sets of their input transitions), and $\URSpls$
models $\mc{U}$.
The task of $\SplitBlock$ is to update these three variables according to the
splitting of a block $C$.
Before calling $\SplitBlock$, the states $q$ that should go to one of the
halves have been marked by calling $\BRP.\Mark(q)$ for each of them.

Line~1 unmarks all states of $C$ and either splits $C$ in $\BRP$ updating
$\mc{B}$, or detects that one of the halves would be empty, so $C$ should not
be split.
In the latter case, line~2 exits the procedure.
The total effect of the call and its preceding calls of $\BRP.\Mark$ is zero
(except that the ordering of the states in $\BRP$ may have changed).

From now on assume that both halves of $C$ are nonempty.
Line~3 makes $b$ the index of the bigger half $B$ and $b'$ the index of the
smaller half $B'$.
Because $C$ is no more a block, for each $a \in \Sigma$, the pairs $(C,a)$
are no more splitters, and must be replaced by $(B,a)$ and $(B',a)$, to the
extent that they are nonempty.
For this purpose, lines~4, 5 and~10 scan $B'$ and line~6 scans the incoming
transitions of the currently scanned state of $B'$.
Line~9 marks, for each $a \in \Sigma$, the transitions that correspond to
$(B',a)$.
Line~7 finds the index of $(C,a)$ in $\TRP$, and line~8 adds it to
$\UTouched$, unless it is there already.
After all input transitions of $B'$ have been scanned, lines~11 and~12
discharge the set of affected splitters $(C,a)$.
Line~13 updates $(C,a)$ to those of $(B,a)$ and $(B',a)$ that are nonempty.

Line~14 corresponds to the updating of $\mc{U}$.
If both $(B,a)$ and $(B',a)$ are nonempty splitters, then the index of
$(B',a)$ is added to $\URSpls$, that is, $(B',a)$ is added to $\mc{U}$.
In this case, $(B,a)$ inherits the index of $(C,a)$ and thus also the presence
or absence in $\mc{U}$.
If $(B,a)$ is empty, then $(B',a)$ inherits the index and $\mc{U}$-status of
$(C,a)$.
If $(B',a)$ is empty, then $(C,a)$ does not enter $\UTouched$ in the first
place.
To summarize, if $(C,a) \in \mc{U}$, then all of its nonempty heirs enter
$\mc{U}$; otherwise only the smaller heir enters $\mc{U}$, and only if it is
nonempty.
This is equivalent to abstract lines 13\ldots14.
Regarding abstract line 15, $(C,a)$ disappears automatically from $\mc{U}$
because its index is re-used.

Lines~15\ldots18 implement the total effect of abstract lines~4\dots5.
The initial value of $\BRP$ corresponds to $\mc{B} = \{Q\}$.
Line~15 makes $\TRP$ contain the sets of input transitions of all nonempty
splitters $(Q,a)$ (where $a \in \Sigma$), and line~16 puts them all to
$\mc{U}$.
If $Q = F$, then lines~17 and~18 have no effect.
Otherwise, they update $\mc{B}$ to $\{F, Q-F\}$, update $\TRP$ accordingly,
and update $\URSpls$ to contain all current nonempty splitters.

Lines~19 and~20 match trivially abstract lines~6 and~7.
They choose some nonempty splitter $(B,a)$ for processing.
Lines~21\ldots26 can be thought of as being executed between abstract lines~7
and~8.
They mark the states in $C_1$ for every $C$ that is scanned by abstract
line~8, and collect the indices of those $C$ into $\BTouched$.
Lines~27 and~28 correspond to abstract line~8, and abstract lines~9\ldots15
are implemented by the call $\SplitBlock(b)$.
Lines~1 and~2 have the same effect as abstract lines~9\ldots11.
Line~3 implements abstract line~12.
The description of line~14 presented above matches abstract lines~13\ldots15.
\qed

\begin{theorem}
Given a PT-DFA all whose states are relevant and that has at least one
final state, and assuming that the transitions that have the same label are
given successively in the input, the algorithm in Figure~\ref{concr-algo} runs
in $O(|\delta| \lg |Q|)$ time and $O(|\delta|)$ memory.
\end{theorem}
\proof
The data structures have been listed in this section and they all consume
$O(|Q|)$ or $O(|\delta|)$ memory.
Their initialization takes $O(|Q|+|\delta|)$ time.
Because all states are relevant, we have $|Q| \leq |\delta|+1$, so $O(|Q|)$
terms are also $O(|\delta|)$.

We have already seen that each individual operation in the algorithm runs in
amortized constant time, except for line~15, which takes $O(|\delta|)$ time.
We also saw towards the end of Section~\ref{Abstr_Algo} that each transition is
used at most $\lg |Q| + 1$ times on line~9 of the abstract algorithm.
This implies that line~25, and thus lines~23\ldots26, are executed at most
$|\delta| (\lg |Q| + 1)$ times.
The same holds for lines~28 and~29, because the number of $\Add$-operations on
$\BTouched$ is obviously the same as $\Remove$-operations.
Because $\TRP$-sets are never empty, lines~20 and~21 are not executed more
often than line~25, and lines~22 and~27 are executed at most twice as many
times as line~25.
Line~19 is executed once more than line~20, and lines~15\ldots18 are executed
once.
Line~16 runs in $O(|\delta|)$ and line~17 in $O(|Q|)$ time.

Lines~1\ldots4 are executed at most once more than line~29.
If $\BRP.\Size(b) \geq \BRP.\Size(b')$ on  line~3, then each of the states
scanned by lines~5 and~10 was marked on line~17 or~25.
Otherwise the number of scanned states is smaller than the number of marked
states.
Therefore, line~10 is executed at most as many times as lines~17 and~25, and
line~5 at most twice as many times.
Whenever lines~7\ldots9 are executed anew (or for the first time) for some
transition, the end state of the transition belongs to a block whose size is
at most half of the size in the previous time (or originally), because the
block was split on line~1 and the smaller half was chosen on line~3.
Therefore, lines~7\ldots9 are executed at most $|\delta| \lg |Q|$ times.
Line~6 is executed as many times as lines~7 and~10 together.
The executions of lines~12\ldots14 are determined by line~8, and of line~11 by
lines~4 and~8.
\qed

\vskip-0.3cm
\section{Sorting Transitions}\label{Sort_Trans}

In $\TRP$, transitions are sorted such that those with the same label are next
to each other.
Transitions are not necessarily in such an order in the input.
Therefore, we must take the resources needed for sorting into account in our
analysis.

Transitions can of course be sorted according to their labels with heapsort in
$O( |\delta| \lg |\delta| )$ time and $O(|\delta|)$ memory.
This is inferior to the time consumption of the rest of the algorithm.
Because the labels need not be in alphabetical order, a suitable ordering can
also be found by putting the transitions into a hash table using their labels
as the keys.
Then nonempty hash lists are sorted and concatenated.
This takes $O(|\delta|)$ time on the average, and $O(|\delta|)$ memory.
However, the worst-case time consumption is still $O(|\delta| \lg |\delta|)$.

\newcommand{\idx}{\var{idx}}
A third possibility runs in $O(|\delta|)$ time even in the worst case, but it
uses $O(|\Sigma|)$ additional memory.
That its time consumption may be smaller than memory consumption arises from
the fact that it uses an array $\idx$ of size $|\Sigma|$ that need not be
initialized at all, not even to all zeros.
It is based on counting the occurrences of each label as in exercise 2.12
of~\cite{AHU74}, and then continuing like counting sort.
The pseudocode is in Figure~\ref{sort-trans}.

\begin{figure}[t]
\begin{ohjelma}\rf
$\TRP.\sets$ := $0$\rt
\tb{for} $t \in \delta$ \tb{do}\rt
\>$a$ := $\var{label}[t]$\pp $i$ := $\idx[a]$\rt
\>\tb{if} $i < 1 \vee i > \TRP.\sets \vee \TRP.\smid[i] \neq a$ \tb{then}\rt
\>\>$i$ := $\TRP.\sets + 1$\pp $\TRP.\sets$ := $i$\rt
\>\>$\idx[a]$ := $i$\pp $\TRP.\smid[i]$ := $a$\pp $\TRP.\send[i]$ := $1$\rt
\>\tb{else} $\TRP.\send[i]$ := $\TRP.\send[i] + 1$\rt
$\TRP.\sfirst[1]$ := $1$\pp $\TRP.\send[1]$ := $\TRP.\send[1] + 1$\pp
$\TRP.\smid[1]$ := $\TRP.\send[1]$\rt
\tb{for} $i$ := $2$ \tb{to} $\TRP.\sets$ \tb{do}\rt
\>$\TRP.\sfirst[i]$ := $\TRP.\send[i-1]$\rt
\>$\TRP.\send[i]$ := $\TRP.\sfirst[i] + \TRP.\send[i]$\pp
  $\TRP.\smid[i]$ := $\TRP.\send[i]$\rt
\tb{for} $t \in \delta$ \tb{do}\rt
\>$i$ := $\idx[\var{label}[t]]$\pp $\ell$ := $\TRP.\smid[i] - 1$\pp
  $\TRP.\smid[i]$ := $\ell$\rt
\>$\TRP.\elems[\ell]$ := $t$\pp
  $\TRP.\loc[t]$ := $\ell$\pp $\TRP.\sidx[t]$ := $i$
\end{ohjelma}
\caption{Initialization of $\TRP$ in $O(|\delta|)$ time and $O(|\Sigma|)$
additional memory}\label{sort-trans}
\end{figure}

\vskip-0.3cm
\section{Measurements and Conclusions}\label{MeaCon}

Table~\ref{Meas} shows some measurements made with our test implementations of
Knuutila's and our algorithm.
They were written in C++ and executed on a PC with Linux and 1 gigabyte of
memory.
No attempt was made to optimise either implementation to the extreme.
The implementation of Knuutila's algorithm completes the transition function
to a full function with a well-known construction.
Namely, it adds a ``sink'' state to which all originally absent transitions
and all transitions starting from itself are directed.

The input DFAs were generated at random.
Because of the difficulty of generating a precise number of transitions
according to the uniform distribution, sometimes the generated number of
transitions was slightly smaller than the desired number.
Furthermore, the DFAs may have unreachable states and/or reachable irrelevant
states that are processed separately by one or both of the algorithms.
Running time depends also on the size of the minimized DFA: the smaller
the result, the less splitting of blocks.
We know that the joint effects of these phenomena were small, because, in all
cases, the numbers of states and transitions of the minimized DFAs were $>$
99.4\,\% of $|Q|$ and $|\delta|$ in the table.
Therefore, instead of trying to avoid the imperfections by fine-tuning the
input (which would be difficult), we always used the first input DFA that our
generator gave for the given parameters.

\begin{table}
\newcommand{\vali}{\hspace*{1.95mm}}%
\caption{Running time measurements.
$|\delta| = p|Q||\Sigma|$, where $p$ is given as $\%$.\newline
A: $|Q| = \vali1\,000$ and $|\Sigma| = 100$.
B: $|Q| = \vali1\,000$ and $|\Sigma| = 1\,000$.\newline
C: $|Q| = 10\,000$ and $|\Sigma| = 100$.
D: $|Q| = 10\,000$ and $|\Sigma| = 1\,000$.
}\label{Meas}\smallskip
\newcommand{\sis}{\hspace*{4mm}}%
\begin{tabular}{llrrrrrr}
  & alg.  & 10\,\%\sis  & 30\,\%\sis  & 50\,\%\sis & 70\,\%\sis
	  & 90\,\%\sis & 100\,\%\sis \\
\hline
A & our   & 0.004 0.005 & 0.013 0.014 & 0.024 0.025 & 0.036 0.037
	  & 0.052 0.060 & 0.061 0.062 \\
  & Knu   & 0.026 0.026 & 0.034 0.035 & 0.040 0.041 & 0.045 0.046
	  & 0.048 0.049 & 0.053 0.054 \\
\hline
B & our   & 0.059 0.061 & 0.277 0.279 & 0.549 0.551 & 0.855 0.865
	  & 1.181 1.211 & 1.330 1.416 \\
  & Knu   & 0.467 0.486 & 0.645 0.651 & 0.785 0.795 & 0.893 0.907
	  & 0.971 0.979 & 1.033 1.040 \\
\hline
C & our   & 0.070 0.071 & 0.296 0.301 & 0.574 0.581 & 0.887 0.893
	  & 1.210 1.229 & 1.424 1.434 \\
  & Knu   & 0.526 0.529 & 0.730 0.734 & 0.901 0.904 & 1.027 1.035
	  & 1.128 1.130 & 1.200 1.202 \\
\hline
D & our   & 1.224 1.238 & 4.038 4.087 & 7.132 7.164 & 10.50 10.57
	  & 14.18 14.34 & 16.41 16.48 \\
  & Knu   & 6.324 6.356 & 8.606 8.705 & 10.46 10.64 & 11.91 11.95
	  & 13.00 13.04 & 13.83 13.89 \\
\end{tabular}
\end{table}

The times given are the fastest and slowest of three measurements, made with
$|F| = \frac{|Q|}{2} + d$, where $d \in \{-1, 0, 1\}$.
They are given in seconds.
The number of transitions $|\delta|$ varies between $10\,\%$ and $100\,\%$ of
$|Q||\Sigma|$.
The times contain the special processing of unreachable and irrelevant states,
but they do not contain the reading of the input DFA from and writing the
result to a file.
With $|Q| = |\Sigma| = 10\,000$, Knuutila's algorithm ran out
of memory, while our algorithm spent about 15~s when $p =
\frac{|\delta|}{|Q||\Sigma|} = 10\,\%$ and 32~s when $p = 20\,\%$.

The superiority of our algorithm when $p$ is small is clear.
That our algorithm loses when $p$ is big may be because it uses both $F$
and $Q-F$ in the initial splitters, whereas Knuutila's algorithm uses only
one of them.
Also Knuutila's algorithm speeds up as $p$ becomes smaller.
Perhaps the reason is that when $p$ is, say, $10\,\%$, the block that contains
the sink state has an unproportioned number of input transitions, causing
blocks to split to a small and big half roughly in the ratio of $10\,\%$ to
$90\,\%$.
Thus small blocks are introduced quickly.
As a consequence, the average size of the splitters that the algorithm uses
during the execution is smaller than when $p = 100\,\%$.
The same phenomenon also affects indirectly our algorithm, probably explaining
why its
running time is not linear in $p$.

Of the three notions of ``smaller'' mentioned in the introduction, our
analysis does not apply to the other two.
It seems that they would require making $\SplitBlock$ somewhat more
complicated.
This is a possible but probably unimportant topic for further work.

A near-future goal of us is to publish a much more complicated, true $O(m \lg
n)$ algorithm for the problem in~\cite{Fer89}, that is, the multi-relational
coarsest partition problem.

\end{document}